\documentclass[manuscript]{aastex63}
\usepackage{amssymb}
\usepackage{textcomp}
\usepackage{tabularx}
\usepackage{longtable}
\usepackage{multirow}
\usepackage{array}
\usepackage{color}
\usepackage{subfigure}
\usepackage{times}
\usepackage{epsfig}
\usepackage{xcolor}
\usepackage{natbib}
\usepackage[T1]{fontenc}
\usepackage{ae,aecompl}

\submitjournal{AJ}

\shorttitle{Kepler Binary Stars in NGC 6819 Open Cluster}
\shortauthors{Soydugan et al.}

\begin{document}

\title{Kepler Binary Stars in NGC 6819 Open Cluster:\\ KIC 5113146 and KIC 5111815}

\correspondingauthor{Esin Soydugan}
\email{esoydugan@comu.edu.tr}

\author[0000-0003-0321-3469]{Esin Soydugan}
\affiliation{\c{C}anakkale Onsekiz Mart University, Faculty of Arts and Sciences, Department of Physics, TR-17020, \c{C}anakkale, Turkey}

\affiliation{\c{C}anakkale Onsekiz Mart University, Astrophysics Research Center and Ulup{\i}nar Observatory, TR-17020, \c{C}anakkale, Turkey}

\author[0000-0002-1972-8400]{Fahri Ali\c{c}avu\c{s}}
\affiliation{\c{C}anakkale Onsekiz Mart University, Faculty of Arts and Sciences, Department of Physics, TR-17020, \c{C}anakkale, Turkey}
\affiliation{\c{C}anakkale Onsekiz Mart University, Astrophysics Research Center and Ulup{\i}nar Observatory, TR-17020, \c{C}anakkale, Turkey}

\author[0000-0002-5141-7645]{Faruk Soydugan}
\affiliation{\c{C}anakkale Onsekiz Mart University, Faculty of Arts and Sciences, Department of Physics, TR-17020, \c{C}anakkale, Turkey}

\affiliation{\c{C}anakkale Onsekiz Mart University, Astrophysics Research Center and Ulup{\i}nar Observatory, TR-17020, \c{C}anakkale, Turkey}

\author[0000-0003-3510-1509]{Sel\c{c}uk Bilir}
\affiliation{Istanbul University, Faculty of Science, Department of Astronomy  and Space Sciences, 34119, Beyaz\i t, Istanbul, Turkey}

\begin{abstract}
In this study, investigation of two double-lined binary stars KIC 5113146 and KIC 5111815 in NGC 6819 open cluster is presented based on both photometric and spectroscopic data. Simultaneous analysis of light and radial velocity curves was made and the absolute parameters of the systems' components were determined for the first time. We find that both systems have F-type main-sequence components. The masses and radii were found to be \emph{M$_{1}$}=1.29$\pm$0.02\emph{M$_{\odot}$}, \emph{R$_{1}$}=1.47$\pm$0.03 \emph{R$_{\odot}$} and \emph{M$_{2}$}=1.19$\pm$0.02 \emph{M$_{\odot}$}, \emph{R$_{2}$}=1.13$\pm$0.02 \emph{R$_{\odot}$} for the primary and secondary components of KIC 5113146; \emph{M$_{1}$}=1.51$\pm$0.08 \emph{M$_{\odot}$}, \emph{R$_{1}$}=2.02$\pm$0.05 \emph{R$_{\odot}$} and \emph{M$_{2}$}=1.19$\pm$0.07 \emph{M$_{\odot}$}, \emph{R$_{2}$}=1.32$\pm$0.04 \emph{R$_{\odot}$} for the components of KIC 5111815, respectively. Evolutionary status of the components was evaluated based on the \texttt{MESA} evolutionary tracks and isochrones. The ages of the KIC 5111815 and KIC 5113146 were derived to be about $2.50\pm0.35$ Gyr and $1.95\pm0.40$ Gyr, respectively. Photometric distances were calculated to be 2850$\pm$185 pc for KIC 5113146 and 3120$\pm$260 pc for KIC 5111815. The results reveal that both KIC 5113146 and KIC 5111815 systems are the most likely member of NGC 6819. 
\end{abstract}


\section{Introduction}
Detached binary stars are important objects for the precise determination of the basic parameters of the stars, the detection of the relationships between fundamental parameters of the stars, as well as the testing of the stellar structure and evolutionary models \citep[][]{Andersen91, Torres10, Eker14, Eker15, Eker18}. On the other hand, a binary star in a stellar cluster search also provides a significant advantage for testing and limiting some astrophysical parameters of both cluster and binary star.

The fact that a star is a cluster member allows us to estimate its age, chemical abundance, and distance by using cluster properties \citep[e.g.][]{Southworth04, Torres19}. If the studied star is also a member of a binary star, its basic parameters, which can be found in the analysis of light and radial velocity variations, can be tested with cluster parameters. However, except for well-detached, double-lined, non-active eclipsing binaries, different properties of binary stars (tidal interactions, mass and angular momentum losses and transfers, magnetic activity, etc.) pose difficulties in determination their astrophysical parameters. In addition, a binary star, as in this study, may not show any eclipse due to its orbital properties and geometry of the components, or only one eclipse can be observed in its light curve (LC). In this case, the determined basic parameters of the components must be tested in other ways, otherwise the uncertainty will be quite large. These types of tests can be done when such systems are cluster members \citep[e.g.][]{Meibom09, Brogaard11, Sandquist13, Brewer16, Hoyman19}.

NGC 6819 ($\alpha_{2000.0}$ = 19$^{\rm h}$ 41$^{\rm m}$ 18$^{\rm s}$.48, $\delta_{2000.0}$= +40$^{\rm o}$ 11$^{\rm '}$ 24$^{\rm ''}$.00; $l=73^{\rm o}.9778$, $b=+08^{\rm o}.4808$), which is a well-known, an intermediate age ($\sim$ 2.5 Gyr) open cluster, firstly, was discovered by Caroline Herschel in 1784 years. By the reason of both the clusters age and favorable location in the Kepler area, this cluster is one of the most appropriate for studying stars have different evolution stages. Therefore, many researchers investigated it to obtain main astrophysical parameters (e.g. reddening, metallicity, distance, and age). Studies on NGC 6819 were mainly made as photometric research \citep[e.g.][]{Kalirai01, Carraro13, Anthony-Twarog14, Ak16}, stellar variability \citep[e.g.][]{Sandquist13, Balona13, Milliman14, Brewer16}, spectroscopic studies \citep{Hole09, Milliman14, Lee-Brown15}, metallicity \citep[e.g.][]{Bragaglia01, Anthony-Twarog14, Ak16, Bossini19}, and age \citep[e.g.][]{Anthony-Twarog14, Bedin15, Ak16, Bossini19}. In addition, colour excess $E(B-V)$ and distance module or distance were calculated by different authors \citep[e.g.][]{Kalirai01, Rodrigues14, Anthony-Twarog14, Bedin15, Ak16, Bossini19}. In order to determined iron and other metal abundance, \citet{Lee-Brown15} analysed high resolution spectra of 333 stars in the cluster around Li 6708~\AA~line and determined the metal abundance of cluster to be ${\rm [Fe/H]}=-0.02\pm0.02$ dex. \citet{Ak16} calculated the metallicity of the NGC 6819 as ${\rm [Fe/H]}=0.051\pm0.020$ dex using the photometric metal abundance calibration of \citet{Karaali11}. The structural parameters of the NGC 6819 was also determined by \citet{Ak16}: the central stellar density (\emph{f$_{0}$} = 13.18$\pm$0.46 stars arcmin$^{-2}$), core radius (\emph{r$_{c}$} = 3.65$\pm$0.38 arcmin), background stellar density (\emph{f$_{bg}$} = 5.98$\pm$0.45 stars arcmin$^{-2}$) and angular radius of the cluster ($r\approx9$ arcmin) were determined. In addition, \citet{Ak16} calculated the colour excess of the cluster as $E(B-V)=0.130\pm0.035$ mag by fitting the Zero Age Main-Sequence (ZAMS) curve of \citet{Sung13} to the positions of most probably cluster member dwarf stars in the $U-B\times B-V$ two-colour diagrams. Since NGC 6819 is in the scanning area of the {\it Kepler} satellite, it was possible to examine the variations of the large number of variable stars in the cluster by taking accurate {\it Kepler} photometric data.

NGC 6819 open cluster was also studied for different types of variable stars using high precision {\it Kepler} data \citep[e.g.][]{Miglio12, Corsaro12, Sandquist13, Wu14}. Studies on binary stars in NGC 6819 were presented by \citet{Sandquist13} for KIC 5024447 and \citet{Jeffries13} for KIC 5113053. The triple-lined system, WOCS 24009 included a short-period detached binary, was investigated by \citet{Brewer16} to obtain absolute parameters of the components and give a constrain to the age of the cluster.

In this paper, two neglected double-lined binary stars KIC 5113146 and KIC 5111815 taken place in NGC 6819 open cluster are studied based on analysis of {\it Kepler} photometric data together with the published radial velocities and evolutionary models. Information about the observational data and its modelling is given in Sections 2 and 3, respectively. Results and their discussions included main astrophysical parameters, evolutionary status of the targets and their cluster membership possibilities are detailed in Section 4. Conclusions with our final remarks are presented in the last section.

\section{Observational Data}

The spectroscopic data used in this study were taken from \citet{Milliman14}, who made radial velocity (RV) measurements of 93 spectroscopic binaries including KIC 5113146 and KIC 5111815 and determined orbital parameters of these systems in NGC 6819. In that study, spectroscopic data was derived from the Hydra Multi Object Spectrograph \citep[MOS,][]{Barden94} on the WIYN 3.5m telescope. \citet{Milliman14} used the TODCOR code to measure RV values of the components of the double-lined spectroscopic binaries. The orbital elements of our targets obtained by \citet{Milliman14}, which were used input parameters for our solutions are listed in Table 1. In the study of \citet{Milliman14}, the root mean squared residual velocities from the orbital solutions for the components of KIC 5113146 and KIC 5111815 were calculated less than 1.2 km s$^{-1}$ and 2.4 km s$^{-1}$, respectively.

Photometric data of the studied binaries was taken from the public archive of {\it Kepler} satellite. As well known, precision light curves of many variable stars were acquired by {\it Kepler} observations, although its main aim is to discover Earth-like exoplanets in the area of Cygnus, Lyra and Draco constellations. Details of {\it Kepler} photometry and observing program were defined by many researchers \citep[e.g.][]{Borucki10, Koch10, Batalha10, Caldwell10, Gilliland10}. {\it Kepler} light curves of the targets were taken from MAST data archive\footnote{https://mast.stsci.edu/portal/Mashup/Clients/Mast/Portal.html}. The errors in photometric data used in this study are in the range of 0.1-0.2\%.

\begin{table}
  \caption{Orbital parameters for KIC 5113146 and KIC 5111815 obtained by \citet{Milliman14}.}
  \begin{center}
           \begin{tabular}{lllll}
      \hline\hline \\[-8pt]
          Parameter    & &    KIC 5113146       &                    & KIC 5111815                     \\
      \cline{1-1}\cline{3-3}\cline{5-5}\\[-8pt]
      \emph{T$_{0}$} (HJD)          & & 2453192.52 $\pm$ 0.05      &  & 2455064.78 $\pm$ 0.24  \\
      \emph{P$_{orb}$ }(day)        & & 18.78911 $\pm$ 0.00020      & &  3.574911 $\pm$ 0.000021 \\
      \emph{V$_{\gamma}$ }(km s$^{-1}$) & & 3.72 $\pm$ 0.21        & &  2.9 $\pm$ 0.3 \\
      \emph{K$_{1}$ } (km s$^{-1}$) & & 56.7 $\pm$ 0.6             & & 76.3 $\pm$ 0.5 \\
      \emph{K$_{2}$ } (km s$^{-1}$) & & 61.4 $\pm$ 0.4             & & 97.3 $\pm$ 0.5 \\
      \emph{e}                      & & 0.397 $\pm$ 0.06           & & 0.010 $\pm$ 0.005 \\
      \emph{w} (deg)                & & 305.8 $\pm$ 1.4            & & 326 $\pm$ 25\\
    \emph{a$_{1}\sin i$} (10$^{6}$\,{\rm km}) &  & 13.45 $\pm$ 0.16      & & 3.75 $\pm$ 0.03 \\
    \emph{a$_{2}\sin i$} (10$^{6}$\,{\rm km}) &  & 14.56 $\pm$ 0.13      &  & 4.78 $\pm$ 0.02 \\
      \emph{M$_{1}\sin^{3}\,i$} (\emph{M$_{\odot}$}) & &  1.29 $\pm$ 0.03 & & 1.09 $\pm$ 0.01 \\
      \emph{M$_{2}\sin^{3}\,i$} (\emph{M$_{\odot}$}) & &  1.19 $\pm$ 0.03 & & 0.85 $\pm$ 0.01 \\
       \hline
     \end{tabular}\\
     \end{center}
\end{table}

\section{Modeling of Binary Stars}

The targets in this study are slightly different from binary systems that are frequently studied in the literature. KIC 5111815 does not show any eclipse, while KIC 5113146 exhibits light variations with a single eclipse. Before analyzing the light changes taking into account binary effects, we first wanted to make a frequency scan in the light curve of KIC 5111815, which does not show any eclipse. The software package Period04 v1.2 \citep{Lenz05} was preferred for the Fourier analysis of the {\it Kepler} light curve of KIC 5111815. The dominant frequency value was obtained to be \emph{f$_{1}$}=0.5601 c d$^{-1}$ with a $S/N=16.9$. The period equivalent of the most prominent frequency is $P\approx1.7854$ days, which is approximately half of the system's orbital period ($P_{orb}\approx3.5749$ days). Several frequency values that follow \emph{f$_{1}$} are also found to be near or half the orbital period. As a result of this frequency scan, it can be said that the cause of light changes was most likely due to the binary effects and also magnetic activity of the component(s). Therefore, in order to estimate the geometric and physical parameters of the components, the light curve of KIC 5111815 can be analyzed taking into account the binary effects.

The simultaneous analysis of the light curves and the components' RV values of two double-lined binaries KIC 5113146 and KIC 5111815 were made for the first time based on {\it Kepler} CCD photometric data and published RV measurements by \citet{Milliman14}. The theoretical light and radial velocity curves related to both systems were calculated with the Wilson Devinney code (W-D) \citep{Wilson71} as it was used in many studies \citep[e.g.][] {Soydugan11, Soydugan13, Tuysuz14, Wolf20, Wang20}. The W-D code for the modeling of RV and LC data of binary stars is considered one of standard applications. It has been tested for many targets and proven reliable. It is suitable well for making analysis of the observational data for different types of binary stars, for instance, interacting binaries, active binaries, non-eclipsing binaries, detached and semi-detached binaries, X-ray binaries, contact binaries etc. \citep[e.g.][] {Ostrov02, Soydugan03, Bozic07, Rozyczka13, Ren17}. The software takes into account some other important effects (e.g. rotational and tidal distortion, the reflection effect, gravity and limb darkening) and utilizes the Roche geometry to approximate the components' shapes when analyzing LC and RV data of binary systems. For these reasons, we decided to use the W-D code to analyze the RV and LC data of KIC 5111815 and KIC 5113146. In addition, we have preferred to use the W-D code combined with the Monte Carlo simulation to determine uncertainties of the adjustable parameters \citep[e.g.][] {Zola04, Zola10}. Although one of the systems does not show an eclipse and one shows a single eclipse, the fact that the components of the systems have RV data provided us an advantage to take the lead in the analysis and determine the parameters of the components for the first time. 

In order to decide real configuration of KIC 5113146 and KIC 5111815, in the first trials assumed that the systems are detached binaries, we followed the dimensionless potential value of the components for both systems. As the potential values concerning the companions of the targets did not reach theirs inner Roche limits, we decided that these systems indicate a detached configuration. Therefore, W-D code with MOD 2, which corresponds to a detached configuration, was used to obtain final photometric results for the studied systems.

In the analysis, some parameters were taken as fixed using the relevant literature, while some others were selected as adjustable during the iterations. The adjustable parameters are the orbital inclination (\emph{i}), surface temperature of the secondary component (\emph{$T_{2}$}), dimensionless potentials of the primary and secondary components (\emph{$\Omega_{1,2}$}), orbital semi-major axis (\emph{a}) radial velocity of the system's mass center ($V_{\gamma}$), phase shift, fractional luminosity of the primary component (\emph{L$_{1}$}) and third light (\emph{l$_{3}$}). Orbit properties for both systems as given in
Table 1 indicate that orbits of the targets are eccentric and so, input values of \emph{e} and \emph{w} were taken from the orbital solutions given by \citet{Milliman14}. These parameters were adopted in Solution I for KIC 5111815 and taken to be free for KIC 5113146. We made two analysis for KIC 5111815 seen in Table 2 as Solution I and Solution II, which refer to the analysis with \emph{e}$\neq$0 and \emph{e}=0, respectively since during the orbit solutions of the system we have reached very close $\chi^2$ values and also KIC 5111815 was found much older than its circularization time as given in Section 4.2. While mass ratio (\emph{q}) was selected as a adjustable for KIC 5113146, fixed \emph{q} value of KIC 5111815 in Solution I from the orbital solution by \citet{Milliman14} to be 0.784 and was used free in Solution II. However, the same mass ratio value ($q=0.784$) was also found in trial of Solution II for KIC 5111815. Surface temperatures of the primary components for KIC 5113146 and KIC 5111815 were taken to be 6241 K \citep{Prsa11} and 6580 K \citep{Huber14}, respectively. The bolometric gravity-darkening coefficient (\emph{g$_{1,2}$}) were fixed to 0.32 for convective envelopes (Lucy 1967), whihe the bolometric albedo values (\emph{A$_{1,2}$}) were set to 0.5 for convective envelopes (Rucinski 1969) during the analysis.

\begin{table*}
\caption{Parameters obtained from analysis of \emph{Kepler} light
curves and published RV data for KIC 5113146 and KIC 5111815.}
    $$
        \begin{tabular}{llll}
    \hline
Parameter                            & KIC 5113146        & \multicolumn{2}{c}{KIC 5111815}   \\
                                     &                    &Solution I         & Solution II   \\
    \hline

\textit{a} (\emph{R$_{\odot}$})     & 40.18(22)           & 13.69(81) &13.69(75)  \\
\textit{V}$_{\gamma}$ (km s$^{-1}$) & 3.72(50)           & 2.84(21)  & 2.69(21)         \\
\textit{i} ($^{\circ}$)              & 85.47(8)          & 63.1(6.0)& 63.4(6.2)    \\
\textit{T}$_{1}$ (K)                 & 6241$^{a}$         & 6580$^{b}$ & 6580$^{b}$     \\
\textit{T}$_{2}$ (K)                 & 6148(30)           & 6280(20)   & 6280(20)     \\
\textit{q}                          & 0.925(2)            & 0.784$^{c}$  & 0.784(8)    \\
$\Omega_{1}$                        & 28.467(457)           & 7.511(235)   & 7.560(228)   \\
$\Omega_{2}$                        & 34.059(589)          &  9.190(520)   & 9.199(540) \\
\emph{e}                            & 0.404(5)            &  0.010$^{c}$ & 0$^{d}$   \\
\emph{w} (deg)                      & 306(3)              &  326$^{c}$   & 90$^{d}$    \\
Phase shift                         & 0.4003(1)           & 0.349(24)    &  0.349(25)   \\
\textit{L}$_{1}$/(\textit{L}$_{1}$+\textit{L}$_{2}$) & 0.642(24)    & 0.740(30) & 0.740(30)      \\
\textit{L}$_{2}$/(\textit{L}$_{1}$+\textit{L}$_{2}$) & 0.358(16)    & 0.260(40)  & 0.260(40)    \\
\textit{l}$_{3}$                    & 0.337(23)      & -                        & -  \\
{\textit{r}}$_{1}$ (mean)                             & 0.0366(6)   & 0.1486(72) & 0.1475(70)    \\
{\textit{r}}$_{2}$ (mean)                             & 0.0282(5)   &  0.0971(85) & 0.0967(80)\\

\noalign{\smallskip} \hline \noalign{\smallskip}
\end{tabular}
    $$
\begin{center}
    \begin{minipage}{120mm}
($^{a}$) \citet{Prsa11}, ($^{b}$) \citet{Huber14}, ($^{c}$) \citet{Milliman14}, ($^{d}$) Assumed
\end{minipage}
\end{center}
\end{table*}

The analysis were ended up that the corrections related to adjustable parameters became smaller than the corresponding probable errors. The final parameters gained from simultaneous solutions of the radial velocity and {\it Kepler} light curves for KIC 5113146 and KIC 5111815 binaries are given in Table 2 and agreement of the observational data with theoretical light and radial velocity curves are shown in Figure 1 and Figure 2, respectively. As seen in the light curve of KIC 5113146, only one minimum appears. The components of this system are quite distant from each other (\textit{a}$\approx 40.2\pm0.2 \emph{R$_{\odot}$}$) and the mean fractional radii of the components were calculated as 0.0336 and 0.0279 for the primary and secondary components, respectively. In addition, a third light, which is estimated to contribute about 34\% of the total light of KIC 5113146, is one of the important findings to be tested. The shallow minimum and precisely determined mass ratio led to the conclusion that the light curve can only be represented by such a third light in the system. As seen in the Figure 1, the light curve of KIC 5111815 does not show any eclipse, but it is estimated that there are variations in the light curve caused by binary effects and a possible magnetic activity of the component(s). Figure 3 shows a comparison between observed light curve and theoretical ones with different orbital inclination values for KIC 5111815. The beginning of the eclipse at approximately \textit{i}=75$^{\circ}$ can be clearly traced from the figure. Besides, it is seen that non-eclipse variations occur at different amplitudes depending on the orbital inclination. As seen in the Figure 4, which indicates geometric configurations of the systems observed at 0.5 and 1.0 orbital phases, only one eclipse occurs in the light curve of KIC 5113146, while KIC 5111815 has no any eclipse.

\begin{figure}
\begin{center}
\includegraphics*[scale=.65,angle=000]{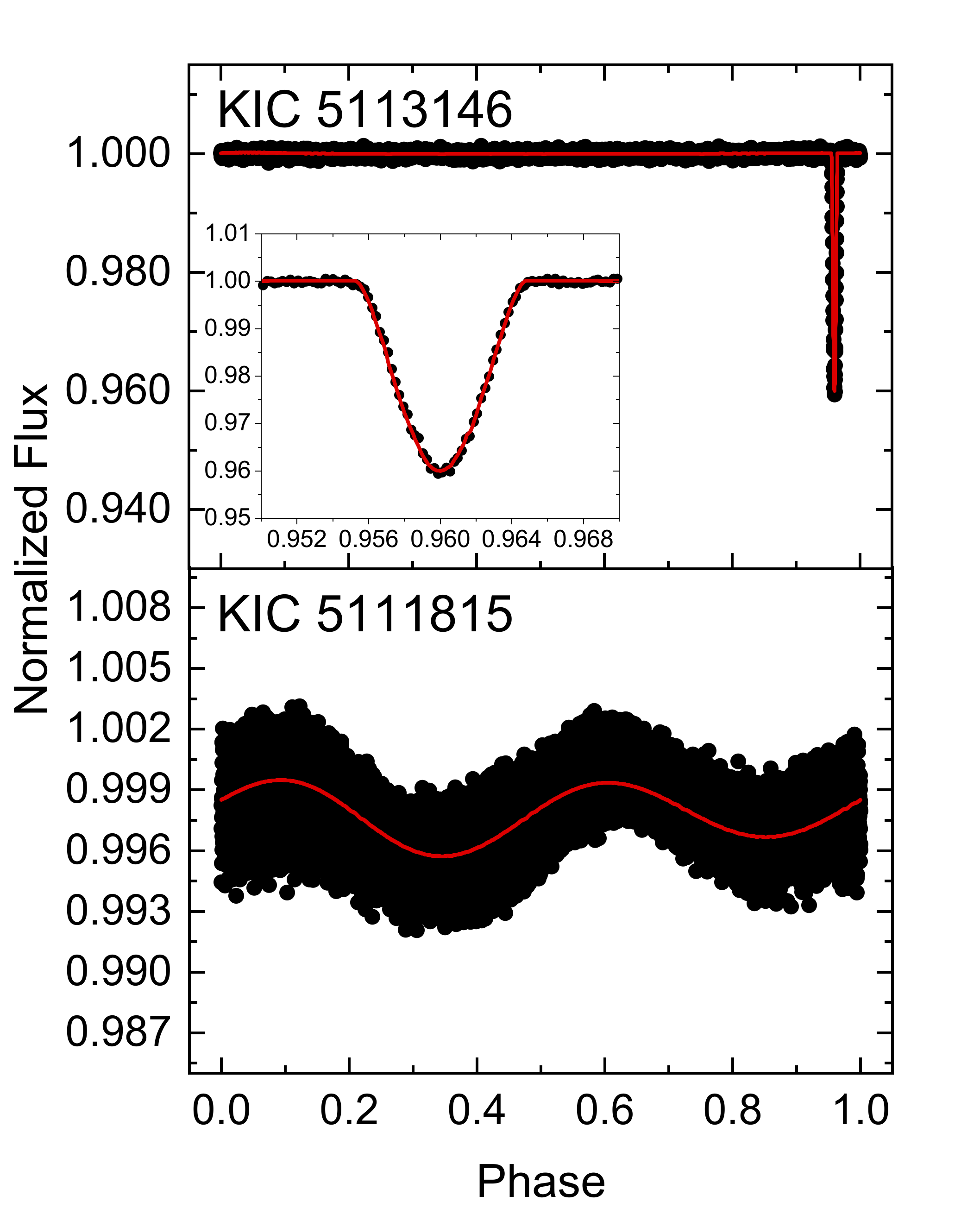}
\caption{\emph{Kepler} light curves and their theoretical representations for KIC
5113146 and KIC 5111815.} \label{fig1}
\end{center}
\end{figure}

\begin{figure}
\begin{center}
\includegraphics*[scale=0.65,angle=000]{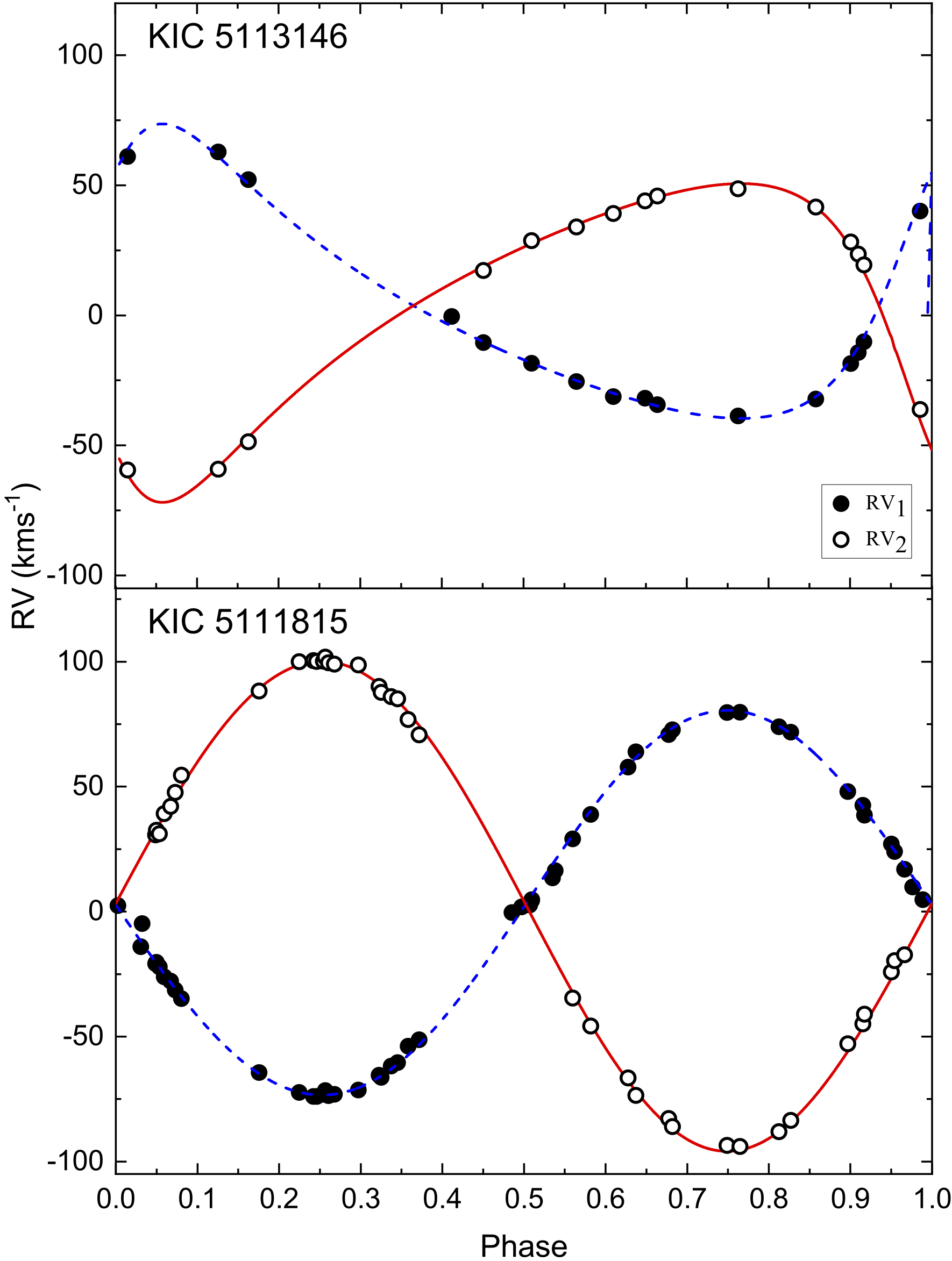}
\caption{RV values of the primary (RV$_{1}$) and secondary (RV$_{2}$) components of KIC 5113146 and KIC 5111815 published by \citet{Milliman14} together with theoretical curves calculated in this study.} 
\end{center}
\end{figure}

\begin{figure}
\begin{center}
\includegraphics*[scale=.55,angle=000]{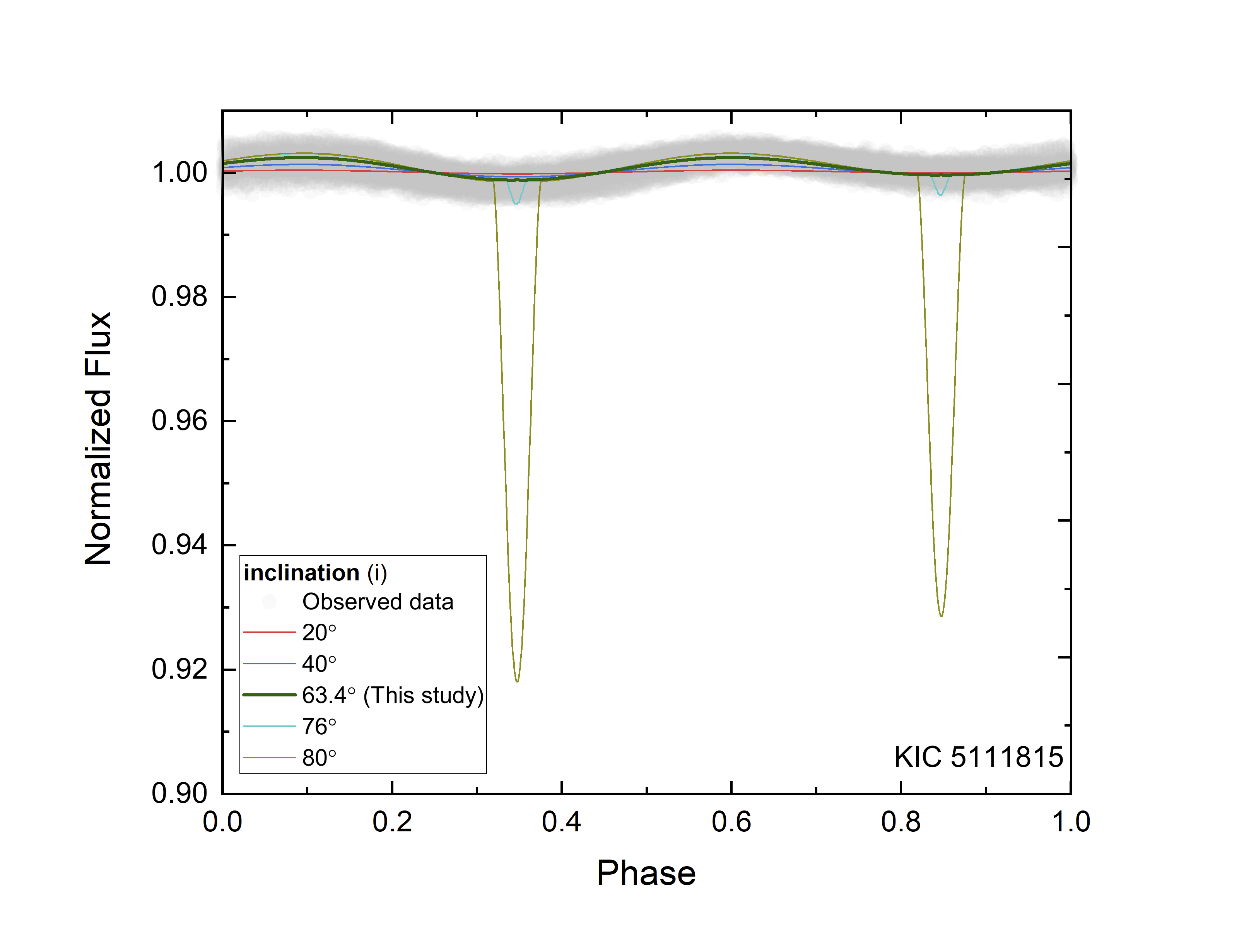}
\caption{Comparison between observational data and theoretical light curves with different orbital inclination values for KIC 5111815.} \label{fig1}
\end{center}
\end{figure}

\begin{figure}
\begin{center}
\includegraphics*[scale=.65,angle=000]{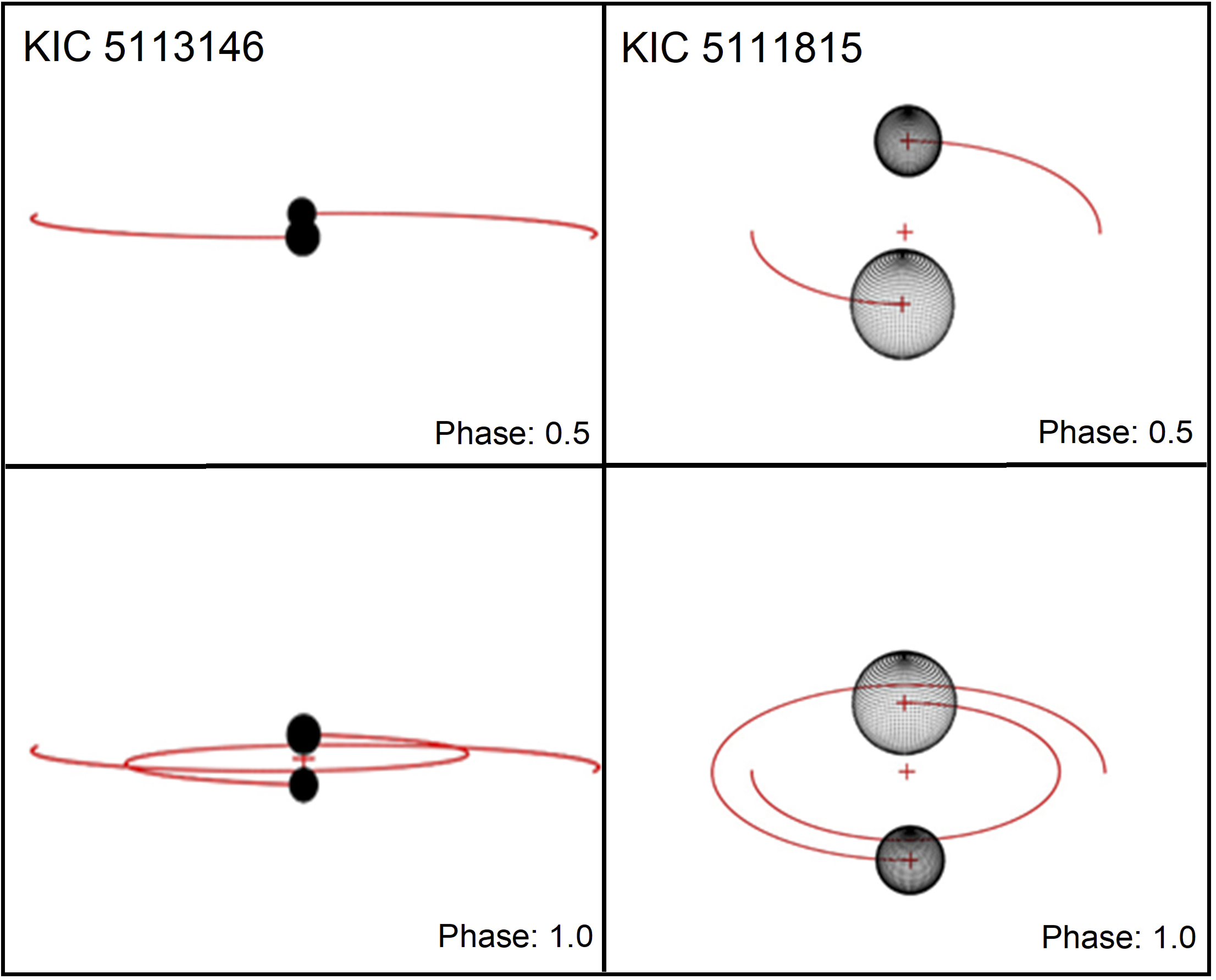}
\caption{Geometric configurations of studied systems at 0.5 and 1.0 orbital phases.} \label{fig1}
\end{center}
\end{figure}

\section{Results and Discussions}

\subsection{Fundamental Astrophysical Parameters}

Modelling {\it Kepler} light curves and also published radial velocities
of the components by \citet{Milliman14} for both KIC 5113146 and
KIC 5111815 provide us to calculate the main-astrophysical
parameters of the systems and their components listed in Table 3.
Although only one eclipse in the light curve of the KIC 5113146 and
no eclipse for KIC 5111815 was observed, the examined systems were
double-lined spectroscopic binaries and also likely to be members of
the open cluster NGC 6819 provided significant advantages when calculating the
basic absolute parameters. We have accepted the results of Solution
II (\emph{e}=0) to calculate the absolute parameters of the
components of KIC 5111815. As seen in Table 3, we calculated the errors are about 6\% for the masses and 2\% for the radii of the components of KIC 5111815. The precision in the masses and radii is about 2\% for the components of KIC 5113146. The bolometric correction values of the components of the targets used in the calculations were taken from \citet{Eker20} with a solar temperature ($T_{eff,\odot}$) of 5772 K and solar bolometric absolute
magnitude ($M_{bol,\odot}$) of 4.74 mag.

Both studied systems are found to be detached binaries. From the
surface potential values of the components and mass ratio of the
systems, filling ratios of the inner Roche volumes of the primary
and secondary companions were calculated to be 23\% and 19\% for KIC
5113146 and 46\% and 38\% for KIC 5111815, respectively. Especially,
KIC 5113146 can be studied as two separate single stars since it is
well-detached system with spherical components and tidal effects can
be neglected.

Updated list of well-known detached binaries with absolute
parameters are presented by \citet{Eker18}. It is clear that in
the appropriate mass ranges, the basic parameters of the targets are compatible with the parameter distributions of the detached binaries. Using the mass values of the components of the
studied systems given in Table 3, the luminosities were calculated
with the help of the mass-luminosity relation (MLR) given by 
\citet{Eker18} for the stars with intermediate mass. We have derived 
the luminosities of the primary and secondary components to be
\emph{$\log$L$_{1}$}=0.49\emph{L$_{\odot}$} and 
\emph{$\log$L$_{2}$}=0.34\emph{L$_{\odot}$} for KIC 5113146 and
\emph{$\log$L$_{1}$}=0.81\emph{L$_{\odot}$} and log
\emph{$\log$L$_{2}$}=0.35\emph{L$_{\odot}$} for KIC 5111815, respectively.
It was found that the luminosity values of KIC 5111815's components
in Table 3, which were calculated by Stefan-Boltzmann law were quite
compatible with those obtained from updated MLR by \citet{Eker18}, 
while there were differences in the those values of KIC
5113146's components in the order of error.

Considering the masses of the components, according to \citet{Eker20}, spectral types of KIC 5113146 and KIC 5111815 can be estimated as F6V+F9V and F2V+F9V, respectively. Photometric distances of the systems were calculated using the colour excess value (\emph{E(B-V)}$\approx$0.13 mag) of the open cluster NGC 6819 given by \citet{Ak16}. As listed in Table 3, the photometric distances of the detached binaries were calculated to be 2850$\pm$185 pc and 3120$\pm$260 pc for KIC 5113146 and KIC 5111815, respectively, which are not far from the \emph{Gaia} DR2 distance values \citep{Gaia18}.

\begin{table*}
  \caption{ Fundamental astrophysical parameters of KIC 5113146 and KIC 5111815.}
  \begin{center}
           \begin{tabular}{lll}
  \hline\hline
     Parameter                                  &    KIC 5113146       & KIC 5111815  \\
  \hline
     \emph{M$_{1}$} (\emph{M$_{\odot}$})        & 1.29$\pm$0.02        & 1.51$\pm$0.08 \\
     \emph{M$_{2}$} (\emph{M$_{\odot}$})        & 1.19$\pm$0.02        & 1.19$\pm$0.07 \\
     \emph{R$_{1}$ }(\emph{R$_{\odot}$})        & 1.47$\pm$0.03        & 2.02$\pm$0.05  \\
     \emph{R$_{2}$ }(\emph{R$_{\odot}$})        & 1.13$\pm$0.02        & 1.32$\pm$0.04  \\
     \emph{ T$_{1}$} (K)                        & 6241$\pm$200$^{a}$   & 6580$\pm$200$^{d}$  \\
     \emph{ T$_{2}$} (K)                        & 6148$\pm$230         & 6280$\pm$235  \\
     log \emph{L$_{1}$} (\emph{L$_{\odot}$})    & 0.47$\pm$0.06        & 0.84$\pm$0.08 \\
     log \emph{L$_{2}$} (\emph{L$_{\odot}$})    & 0.22$\pm$0.07        & 0.39$\pm$0.09 \\
     log \emph{g$_{1}$} (cgs)                   & 4.21$\pm$0.01        & 4.00$\pm$0.03 \\
     log \emph{g$_{2}$} (cgs)                   & 4.40$\pm$0.02        & 4.27$\pm$0.04 \\
     \emph{M$_{bol,1}$} (mag)                   & 3.57$\pm$0.14        & 2.65$\pm$0.21 \\
     \emph{M$_{bol,2}$} (mag)                   & 4.20$\pm$0.17        & 3.76$\pm$0.23 \\
     \emph{BC$_{1}$} (mag)                      & 0.07$^{b}$           & 0.09$^{b}$  \\
     \emph{BC$_{2}$} (mag)                      & 0.05$^{b}$           & 0.07$^{b}$  \\
     \emph{M$_{V,1}$} (mag)                     & 3.50$\pm$0.14        & 2.56$\pm$0.21 \\
     \emph{M$_{V,2}$} (mag)                     & 4.15$\pm$0.17        & 3.69$\pm$0.24 \\
     \emph{V} (mag)                             & 15.257$\pm$0.003$^{e}$& 15.079$\pm$0.003$^{e}$   \\
     \emph{E(B-V)} (mag)                        & 0.13$^{c}$           & 0.13$^{c}$    \\
     Spectral types                             & F6V+F9V              & F2V+F9V \\
     Orbital separation (\emph{R$_{\odot}$})    & 40.18$\pm$0.22       & 13.69$\pm$0.09 \\
     ($V_{sync}\sin i)_{1}$ (km s$^{-1}$)       & 3.96$\pm$0.07        & 28.9$\pm$0.6 \\
     ($V_{sync}\sin i)_{2}$ (km s$^{-1}$)       & 3.05$\pm$0.06        & 18.9$\pm$0.4 \\
     Photometric distance (pc)                  & 2850$\pm$185         & 3120$\pm$260 \\
     \emph{Gaia}-DR2 distance (pc)              & 2761$\pm$168         & 3581$\pm$313  \\
      BJ (2018) distance (pc)                   & $2562^{+156}_{-140}$ & $3252^{+282}_{-242}$ \\
     Age (Gyr) MESA model                       & 2.50$\pm$0.35        & 1.95$\pm$0.40 \\
      Age (Gyr) PARSEC isochrone               & 2.70$\pm$0.67  & 1.80$\pm$0.16 \\
    \hline
          \end{tabular}\\
      \footnotesize ($^{a}$) \citet{Prsa11}, ($^{b}$) \citet{Eker20}, ($^{c}$) \citet{Ak16}, ($^{d}$) \citet{Huber14}, ($^{e}$) \citet{Milliman14}
     \end{center}
    \end{table*}
    
\newpage
\subsection{Evolutionary Status}

Main astrophysical parameters, especially surface gravity values (log \emph{g}),
give some clues for the evolution of the systems' components. The log \emph{g}  values calculated in this study and presented in Table 3 reveal that both components of the target binaries are main-sequence stars. Unlike other component stars, the primary component of KIC 5111815 has a
slightly smaller log \emph{g} value than those of main-sequence
stars with the same mass according to \citet{Eker20}, suggesting
that it is approaching the Terminal Age Main Sequence (TAMS) line.

In order to better understand the orbital dynamics of binary systems
from the moment of their formation to the present day and the
evolutionary states, binary star evolution models are very useful.
To do more detailed research on the evolution of KIC 5113146 and KIC
5111815, we produced binary star models using MESA (Module for
Experiments in Stellar Astrophysics) code with the version of 8845 \citep{Paxton11, Paxton13, Paxton15, Paxton18}. Applications of \texttt{MESA} binary module
can be reviewed in different studies, which also includes details
on model calculations \citep[e.g.][]{Paxton15, Streamer18, Rosales19, Soydugan20}.

In this study, magnetic braking effect is taken into consideration
in the models and the convective approach of \citet{Hut81} is used for
circularization and synchronization considering the envelope
structures of the stars. During the calculations, the metallicity
value of the systems was also changed between Z=0.012 and Z=0.030
with intervals of 0.001 since the metallicity of NGC 6819 was
reported to be Z=0.016 by \citet{Ak16}. While producing
evolution models, the mass values of the components of the targets
calculated in this study were used as the initial values since the
targets are detached binaries. As a result of the obtained models,
the evolution start period and initial eccentricity of the orbit for
KIC 5113146 were found to be 18.80 days and \emph{e}=0.404,
respectively, which are very close to the current values. For KIC
5111815, the initial value of the orbital period was 1.95 days, while
the initial eccentricity was found to be \emph{e}=0.09. From
the evolution models in different metal abundances, we
estimated the metal abundance values Z=0.022 and Z=0.019 for KIC
5113146 and KIC 5111815, respectively, which indicates that the systems
are a little metal rich than the cluster's metallicity value
(Z=0.016) given by \citet{Ak16}. However, these values need to
be tested with high resolution spectroscopic data.

In the diagram of $\log L\times \log T_{eff}$ or Figure 5,
the locations of the studied system's components, the
evolutionary tracks from the \texttt{MESA} binary module with
metallicity values of Z=0.020 and Z=0.024, the isochrones curve
for the age of NGC 6819 (2.4 Gyr) and also some of possible members 
of the cluster can be seen. In order to estimate ages of the studied systems,
we have tested different isochronous curves between 1.5 Gyr and 
3 Gyr. As a result of this, the ages were found to be 2.50$\pm$0.35 Gyr and 1.95$\pm$0.40 Gyr for KIC 5113146 and KIC 5111815, respectively.

In the study, apart from the \texttt{MESA} evolution models \citep{Paxton11}, the ages of the two systems were also tested with \texttt{PARSEC} isochrones \citep{Bressan12}. Since these two systems are classified as a detached binary stars in this study, they have not yet interacted with each other in terms of binary star evolution. Therefore, the ages of the component stars in the target systems were also determined by \texttt{PARSEC} isochrone. The stellar age probability density function calculated using Bayesian statistics to estimate a precise age with \texttt{PARSEC} isochrone matching. This procedure is based on Bayesian statistics which includes $G(\tau)$ probability density functions  that are obtained from comparative calculations of the theoretical model parameters and the observational parameters \citep{Pont04, Jorgensen05, Duran13, Onal18, Karaali19}. In this method, the age corresponding to the maximum $G(\tau)$ value in the distribution of the posterior probability density function constituted by age is considered as the most likely age of the star. Details of age estimation with Bayesian approach are described by \citet{Onal18} and \citet{Sahin20}.

\begin{figure*}[!ht]
\begin{center}
\includegraphics[scale=0.53]{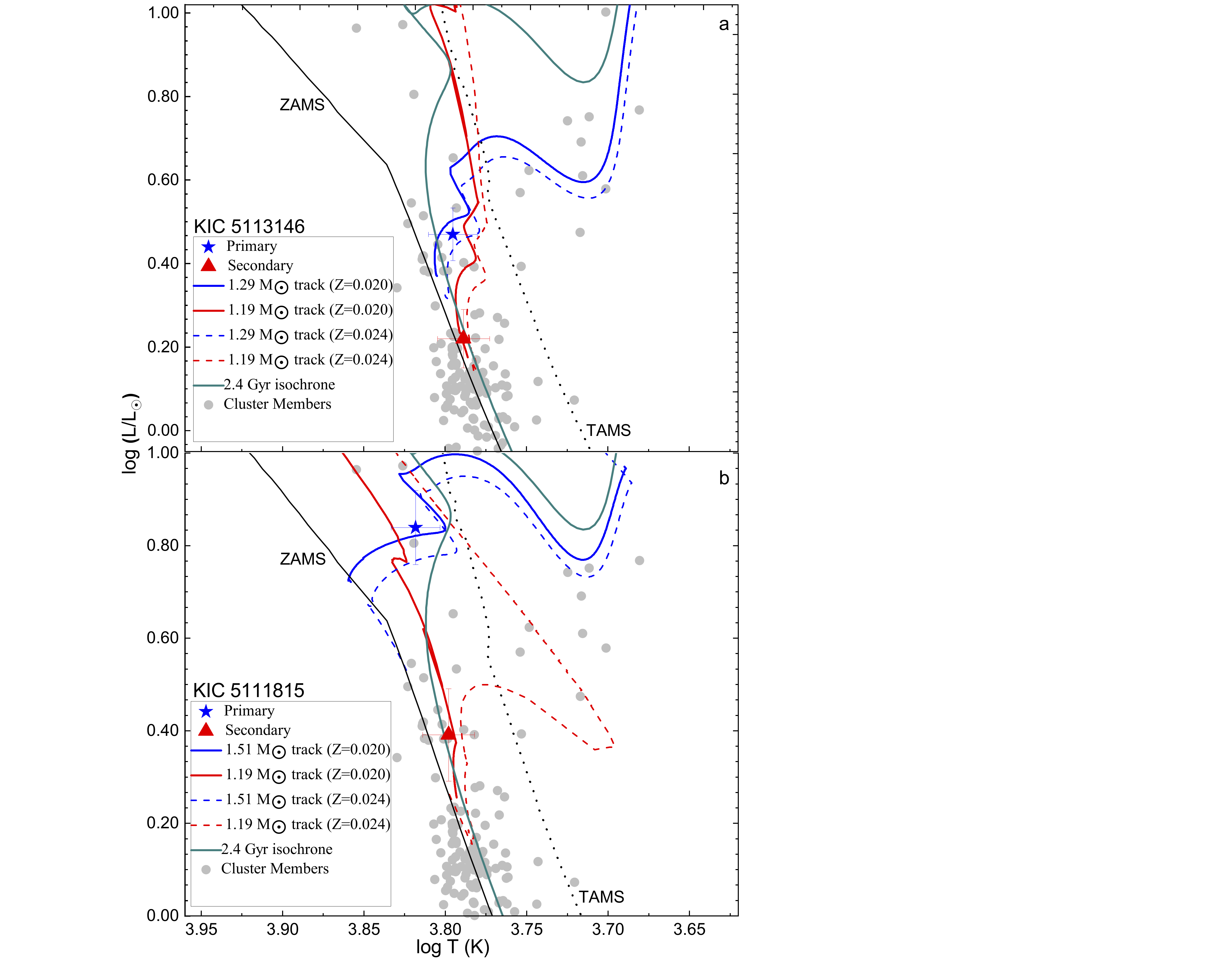}
\caption{Positions of the components of KIC 5113146 (a) and KIC 5111815 (b) shown by filled star (primary) and triangle (secondary) in plane of $\log L \times \log T_{eff}$. Dotted lines refer to Terminal Age Main Sequence (TAMS), while continuous line indicate ZAMS. Evolutionary tracks for calculated masses according to metallicities Z=0.020 (continuous blue line) and Z=0.024 (dashed red line) are shown for both panels. Filled circles indicate possible members of NGC 6819 \citep{Ak16}. Isochronous curve for the age of 2.4 Gyr of the cluster is also shown with continuous green line. Theoretical curves in both panels were calculated by \texttt{MESA} code.} 
\label{fig5}
\end{center}
\end{figure*}

In order to indicate the results of age estimation in this study, posterior probability distributions as a function of age for four component stars of typical ages are shown in Figure 6. The ages corresponding to the maximum $G(\tau)$ values of the primary and secondary components of the KIC 5113146 and KIC 5111815 systems were calculated as (2.85$\pm$0.80, 2.35$\pm$1.25) Gyr and (1.80$\pm$0.17, 1.85$\pm$0.67) Gyr, respectively. The weighted mean ages of the KIC 5113146 and KIC 5111815 are 2.70$\pm$0.67 Gyr and 1.80$\pm$0.16 Gyr, respectively, as well. The position of each star on the $\log g \times T_{eff}$ diagram and the most likely \texttt{PARSEC} isochrone, fit by the Bayesian method, are shown in the top panels of Figure 6. When the $G(\tau)$ distribution of the component stars is analyzed, it can be seen that the error in the ages of the primary components is smaller than of secondary components. Since the mass values of the primaries ($q<1$) are bigger than secondary's ones, the primary components are in a further evolutionary phase within the main sequence band. Due to these characteristics of the primary companions, the ages of these stars are calculated more precise than their secondary stars. Therefore, the relative age errors of the secondary components are higher than that of the primary components. As a result, it has been determined that the star ages calculated from the \texttt{PARSEC} and \texttt{MESA} evolution codes are quite compatible within the age errors. Moreover, the ages calculated for the component stars were compared with those given in the studies in the literature. Considering the study of \citet{Reinhold15}, which determines the gyrochronology ages of 17623 stars in the {\it Kepler} star field, it was determined that only the age of KIC 5111815 in our sample was determined. \citet{Reinhold15} gives the age value $t=1.62\pm0.86$ Gyr for the star KIC 5111815, and this value is quite compatible with $t=1.80\pm0.16$ Gyr value calculated in our study.

In addition, using the formulation of \citet{Zahn77}, we calculated the
circularization and synchronization timescales of both systems to be
log \emph{t$_{syn}$} = 9.13 years and log \emph{t$_{circ}$} = 12.27
years for KIC 5113146 and log \emph{t$_{syn}$} = 6.33 years and log
\emph{t$_{circ}$} = 8.38 years for KIC 5111815. Comparing observational ages and these timescales ($t_{syn}$ and $t_{circ}$), the components are expected to have been synchronized with their orbit. From this calculations, it can be noted that the orbit of KIC 5113146 will remain eccentric during its evolution. However, the age of KIC 5111815 is much larger than the circularization time scale and so, it can be predicted that its orbit turned into a circle long ago.

\begin{figure}[!ht]
\begin{center}
\includegraphics[scale=0.142]{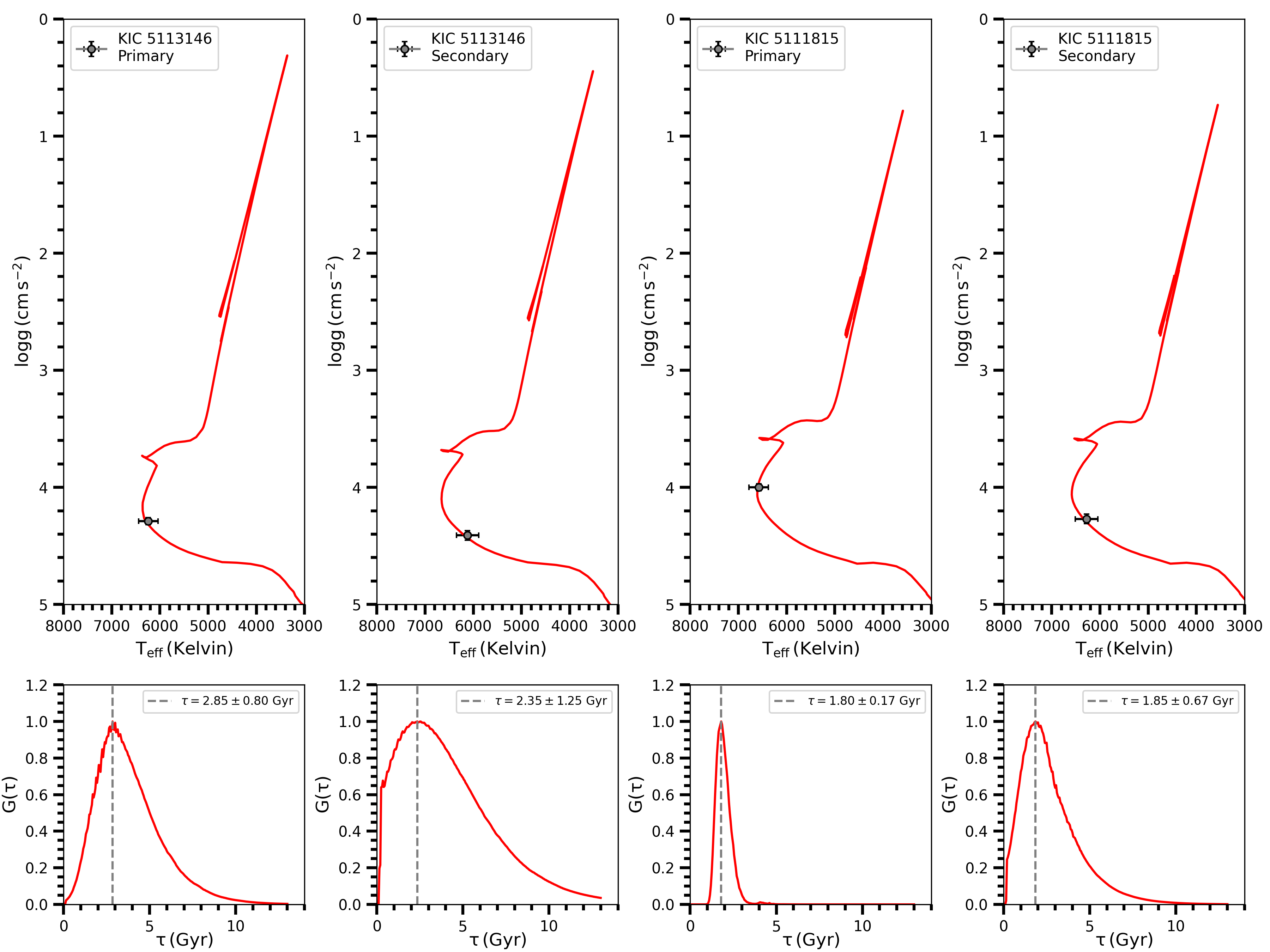}
\caption{$G(\tau)$ posterior probability distribution of stellar ages by Bayesian approach for KIC 5113146 and KIC 5111815 binary systems (lower panel) and the most likely ages of component stars in two systems on $\log g \times T_{eff}$ diagrams (upper panel). Red solid lines and  grey dashed lines represent \texttt{PARSEC} isochrones and most likely age for component stars, respectively.} 
\label{fig7}
\end{center}
\end{figure}

\begin{figure}[!ht]
\begin{center}
5\includegraphics[scale=0.9]{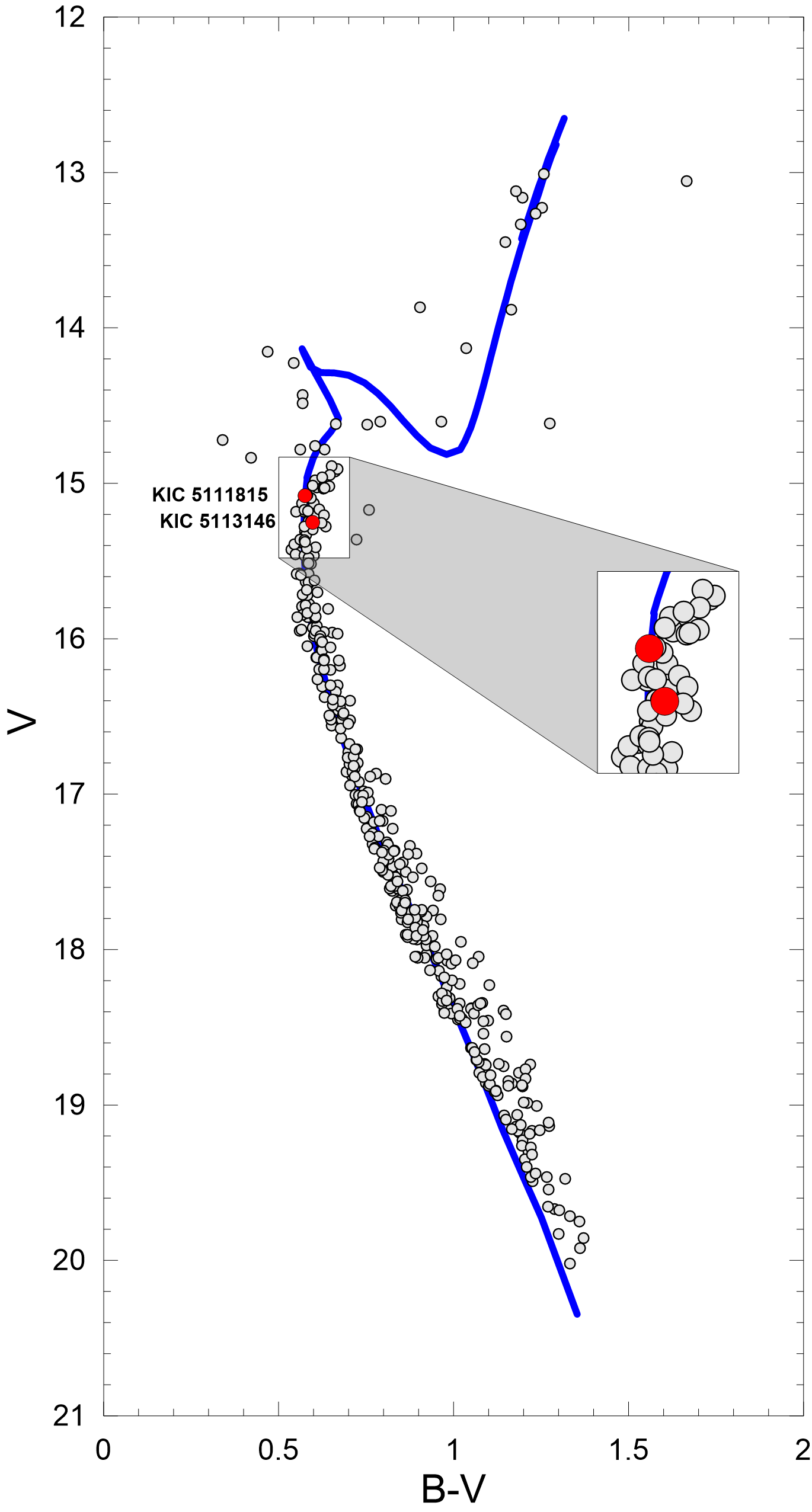}
\caption{Positions of KIC 5113146 and KIC 5111815 binary stars in colour-magnitude diagram are shown by red filled circles together with the most possible cluster members ($P\geq 50\%$) of NGC 6819 are taken from \citet{Ak16}. Continuous line indicates the PARSEC isochrone with estimated cluster age of 2.4 Gyr by \citet{Ak16}.} 
\label{fig7}
\end{center}
\end{figure}

\subsection{Cluster Membership}
The binary stars in this study are positioned similarly to other possible members of NGC 6819 in the colour-magnitude diagram of the cluster as seen in Figure 7. However, the cluster membership status of these systems can also be discussed through different parameters. It is important to use astrometric data in addition to radial velocities in determining the membership of stars to a cluster. Knowing the mean proper motion, radial velocity and distance of the clusters is used to determine the cluster membership of the stars. In our study, radial velocities of the center of mass for KIC 5113146 and KIC 5111815 were taken from \citet{Milliman14}. Moreover, the proper motions, trigonometric parallaxes and radial velocities of these systems were matched with {\it Gaia} DR2 and listed in Table 4. Also, the angular distances of the two systems from the centre of the NGC 6819 are given in the table.

\begin{table*}[!ht]
\setlength{\tabcolsep}{3pt}
  \caption{Astrometric and spectroscopic data of KIC 5113146, KIC 5111815 and NGC 6819. Equatorial coordinates ($\alpha, \delta$), proper motion components ($\mu_{\alpha}\cos\delta$, $\mu_{\delta}$), trigonometric parallaxes ($\varpi$) are taken from {\it Gaia} DR2 \citep{Gaia18}, systematic velocities of the binary systems are calculated in this study while the mean radial velocity of the NGC 6819 is taken from \citet{Soubiran18}. Equatorial coordinates, proper motions and trigonometric parallax data of NGC 6819 open cluster are taken from \citet{Cantat-Gaudin18}. $r$ shows the distance from the cluster center in arcmin for two systems.}
  \label{tab:par}
  \begin{center}
           \begin{tabular}{cccccccc}
  \hline\hline
    Object       & $\alpha_{J2000}$  & $\delta_{J2000}$ & $\mu_{\alpha}\cos\delta$ & $\mu_{\delta}$   & $\varpi$         & $V_{\gamma}$  & $r$      \\
                  & (hh:mm:ss.ss)     & (dd:mm:ss.ss)    & (mas yr$^{-1}$)          & (mas yr$^{-1}$)  & (mas)            & (km s$^{-1}$) & (arcmin)\\
  \hline
KIC 5113146       & 19 41 37.68       & +40 14 32.61     & -3.111$\pm$0.037         & -3.749$\pm$0.042 & 0.362$\pm$0.022  & 3.72$\pm$0.50 & 5.4       \\
KIC 5111815       & 19 40 42.63       & +40 17 08.48     & -2.874$\pm$0.045         & -3.499$\pm$0.043 & 0.279$\pm$0.024  & 2.84$\pm$0.21 & 9.0       \\
\hline
NGC 6819          & 19 41 18.48       & +40 11 24.00     & -2.916$\pm$0.128         & -3.856$\pm$0.140 & 0.356$\pm$0.047  & 3.31$\pm$0.28 &   ---     \\
  \hline
          \end{tabular}\\
      \end{center}
    \end{table*}

\citet{Cantat-Gaudin18} and \citet{Soubiran18} analysing the {\it Gaia} DR2 data of NGC 6819's member stars, presented the mean values of the astrometric and radial velocity of the NGC 6819. \citet{Cantat-Gaudin18}  was calculated the mean proper motions as $ \mu_{\alpha}\cos \delta = -2.916 \pm 0.128$ mas yr$^{-1}$ and $\mu_{\delta} = -3.856 \pm 0.140$ mas yr$^{-1}$ and distance as $d= 2809\pm371$ pc of NGC 6819 using the most probably 1589 membership stars, while \citet{Soubiran18} determined the average radial velocity of the cluster as $V_{\gamma}=3.31\pm0.28$ km s$^{-1}$ using radial velocity values of 48 cluster member stars.

At the first glance, KIC 5113146 and KIC 5111815 are considered to be members of the NGC 6819 open cluster, and appears on the turn-off of the cluster in $V \times B-V$ colour-magnitude diagram (Figure 7). Further proof of membership comes from the proper motions and systemic velocities of these systems. The proper motion and systemic velocities of KIC 5113146 and KIC 5111815 are quite compatible with the mean values calculated for NGC 6819 \citep{Cantat-Gaudin18, Soubiran18}. The photometric distances calculated for the two binaries are quite compatible with the mean distance value of  the NGC 6819 ($d=2809\pm371$ pc) which are calculated from the trigonometric parallax data of 1589 cluster member stars \citep{Cantat-Gaudin18}. The photometric distances calculated in this study and the distances determined from the trigonometric parallax data in the {\it Gaia} DR2 catalog are compared, it is seen that the distances determined from the two methods are compatible within the errors (see Table 3). \citet[BJ,][]{Bailer-Jones18}, who noticed the bias in the trigonometric parallax of the stars in the {\it Gaia} DR2 data and corrected them, the distances of 1.33 billion stars in the {\it Gaia} DR2 catalog were recalculated by Bayesien method. The photometric distances of the two systems studied in this study were compared to \citet{Bailer-Jones18}'s ones, and the distance differences calculated for KIC 5113146 and KIC 5111815 were determined 288 pc and 132 pc, respectively. It was seen that the photometric distance of KIC 5111815 was more compatible with \citet{Bailer-Jones18}'s distance. Moreover, the considering the angular distances of the two systems from the cluster center, it is seen that the both systems are located within the cluster radius \citep[$\emph{r} \approx 9$ arcmin;][]{Ak16, Kalirai01}. The fact that the photometric, astrometric and spectroscopic data of the systems are quite compatible with the values given for the cluster makes KIC 5113146 and KIC 5111815 the most likely members of NGC 6819.

\section{Conclusions}

In this study, it is aimed to investigate binary systems  KIC 5113146 and KIC 5111815,
which have {\it Kepler} light curves and published radial velocity data, since there are no reported studies on photometric solutions and absolute parameters of the systems' components. Considering the calculated masses and surface gravity values of the companions, it is understood that both binary systems consist of F-type main-sequence stars. The mass ratios of the systems and percentage of filling the internal Roche volumes of their components reveal that they are detached systems and exhibit properties compatible with well-known detached binaries with similar masses given by \citet{Eker18}.

Derived mass and radii values have provided us to estimate ages by using \texttt{MESA} evolution model as 2.50$\pm$0.35 Gyr and 1.95$\pm$0.40 Gyr for KIC 5113146 and KIC 5111815, respectively. The fact that the components of the binary systems in this study were not in the advanced evolutionary stages gave the comparison of the ages calculated from the \texttt{MESA} binary evolution models with the \texttt{PARSEC} single evolution model ages. The age values calculated from both models are compatible with each other. Also it can be stated that these values are compatible with the ages of cluster member binary stars and the cluster itself in different studies within the error limits \citep[e.g.][]{Jeffries13, Brewer16, Ak16, Bossini19}.

Evolutionary calculations and the results from the analysis of observational data for KIC 5113146 reveal that its orbital properties have not changed much because of the beginning of its evolution. It can be predicted that the components of the binary star with very smooth spherical geometry evolved like a single star independently. The primary component of KIC 5111815 is located closer to TAMS, while other components of the systems appear to be near ZAMS line in the diagram of $\log L \times \log T_{eff}$.

The proper motion data and radial velocity values of center of mass for KIC 5113146 and KIC 5111815 were compatible with the NGC 6819. The distance values determined in this study related to KIC 5113146 and KIC 5111815 systems were compared with the distance of NGC 6819 cluster given by \citet{Cantat-Gaudin18}. In addition, the distances of \citet{Bailer-Jones18}, which take into account biases in the trigonometric parallax data in {\it Gaia} DR2, are quite compatible with the distances of the systems. Moreover, the angular distances of the two systems to the center of the NGC 6819 confirm the membership of the cluster.

Finally, we note that the results obtained from investigation of F-type {\it Kepler} binaries KIC 511316 and KIC 5111815 can be evaluated in advanced research of cluster and cluster member binary stars. However, high-resolution spectroscopic data is needed to detail the results obtained and test some of them. In particular, research can be carried out on the precise determination of the atmospheric properties of the systems' components and the investigation of the possible third component for KIC 5113146.

\section*{Acknowledgments}
This study was supported by \c{C}anakkale Onsekiz Mart University the Scientific Research Coordination Unit, Project number: FBA-2016-858. The authors are grateful to the anonymous referee for valuable comments and suggestions that helped us to improve the study. We would like to thank Olcay Plevne for his help. This research made use of VizieR and Simbad databases at CDS, Strasbourg, France. This work has made use of data from the European Space Agency (ESA) mission \emph{Gaia} (https://www.cosmos.esa.int/gaia), processed by the \emph{Gaia} Data Processing and Analysis Consortium (DPAC, https://www.cosmos.esa.int/web/gaia/dpac/consortium). Funding for the DPAC has been provided by national institutions, in particular the institutions participating in the \emph{Gaia} Multilateral Agreement.

\end{document}